\documentclass[twocolumn,showpacs,preprintnumbers,amsmath,amssymb]{revtex4}

\usepackage{graphicx}
\usepackage{dcolumn}
\usepackage{bm}
\usepackage{graphics}
\usepackage{amsmath}
\usepackage{amssymb}
\usepackage{amscd}
\usepackage{bbm}
\usepackage{afterpage}
\usepackage{float,times}
\usepackage{subfigure}
\usepackage{rotating}
\usepackage{multirow}
\usepackage{fancyhdr}
\usepackage{epsfig}
\usepackage{theorem}
\usepackage{moreverb}
\usepackage{euscript}
\usepackage{hyperref}

\begin{document}

\newcommand{\veps}{\varepsilon}
\newcommand{\lrarrow}{\leftrightarrow}
\newcommand{\beq}{\begin{equation}}
\newcommand{\eeq}{\end{equation}}
\newcommand{\bea}{\begin{eqnarray}}
\newcommand{\eea}{\end{eqnarray}}
\def\slash{\not\!}
\newcommand{\ud}{\textrm{d}}
\title{Loss of Spin Entanglement For Accelerated Electrons in Electric and Magnetic Fields}
\author{Jason Doukas}
  \email[Email: ]{jadoukas"at"unimelb.edu.au}
   \affiliation{School of Physics,\\
    University of Melbourne, Parkville, Victoria 3010, Australia.}
\author{Lloyd C.L. Hollenberg }
   \affiliation{Centre for Quantum Computer Technology, School of Physics,\\
    University of Melbourne, Parkville, Victoria 3010, Australia.}
\begin{abstract}
Using an open quantum system we calculate the time dependence of the concurrence between two maximally entangled electron spins with one accelerated
uniformly in the presence of constant electric and magnetic fields and the other at rest and isolated from fields. We find at high Rindler temperature the proper time for the entanglement to be extinguished is proportional to the inverse of the acceleration cubed. 
\end{abstract}
\date{\today}
\pacs{03.65.Yz, 03.30.+p,04.70.Dy} \maketitle
\section{Introduction}\label{sec:intro}
Since the discovery of Hawking radiation \cite{Hawking:1974sw} in the early seventies much research into quantum field theory on general classical backgrounds has taken place. One of the most well established results of this field is that of the \textit{Unruh effect} \cite{Fulling:1972md,Unruh:1976db,Davies:1974th} which states that a constantly accelerated particle detector in the Minkowski vacuum will register a thermal bath of particles at a temperature given by:
\beq\label{eqn:rindtemp}
kT=\frac{\hbar a}{2\pi c},
\eeq
where $a$ is the acceleration of the detector.\\

Roughly one decade after the Hawking result, Bell and Leinaas took these observations and applied them to linearly and circularly moving electrons \cite{Bell:1982qr} addressing the anomalous spin depolarization that was observed in storage rings (see also \cite{Akhmedov:2006nd}).\\

Mainly due to the drive for quantum computing, the last twenty years has also seen the rise of quantum information theory, at the heart of which lies the resource of entanglement. Entanglement is a phenomenon unique to quantum systems. Two systems are said to be entangled if they are described by a single state vector that cannot be written as a product of state vectors for each part. The entanglement between two qubits can be quantified by a function called \textit{concurrence} \cite{Wootters:1997id}, taking values on the interval [0,1], where a concurrence of zero (one) represents no (maximal) entanglement. \\

It has only recently been shown that the resource of entanglement is frame dependent \cite{PhysRevLett.89.270402}, in the sense that the distribution of spin and momentum entanglement between two electrons can depend on the speed of the observer measuring these properties. Furthermore, it was shown by \cite{alsing-2003, alsing-2003-91} that the amount of entanglement shared by a pair of field modes in a constantly accelerating cavity entangled with another pair at rest decreases with acceleration. For further developments in this field see \cite{fuentesschuller-2005-95, Alsing:2006cj, cai-2007-76}.\\

While the entanglement between electrons under Lorentz transformations and the entanglement between Dirac field modes in accelerating cavities have both been investigated, a systematic study on the spin entanglement between an accelerating electron and one at rest remains to be done. At first sight there are two effects that require consideration:\\

The first effect is a rotation of the accelerating electron's spin as viewed from the stationary electron frame. Uniform acceleration consists of a continuous sequence of infinitesimal boosts, $d\vec{v}'=\vec{a}d\tau$, where the acceleration $\vec{a}$ and change in velocity $d\vec{v}'$ are measured in the Instantaneous Rest Frame (IRF), i.e., the frame that is momentarily at rest with the moving electron at proper time $\tau$. Viewed from the unaccelerated frame the change in motion comprises of a Lorentz boost $A(\vec{v})$ into the IRF followed by an infinitesimal boost $A(d\vec{v}')$. It is well known that if two successive non-collinear Lorentz boosts take $X$ to $X'$ then the single Lorentz transformation relating these frames is not a pure boost, but rather is the product of a boost and a rotation. The rotation induces an operation on the quantum state called a Wigner rotation. As discussed in \cite{Milburn,PhysRevLett.89.270402} when the particle is not in a momentum eigenstate these Wigner rotations transform the entanglement between the spin and momentum degrees of freedom. \\

The second effect arises from interactions with the thermal environment (\ref{eqn:rindtemp}). As shown in \cite{Bell:1982qr}, the spin of the accelerating electron will flip in response to the Rindler radiation. To understand what effect this has on the spin entanglement it is instructive to consider the simpler spin flip interaction:
\begin{equation}
H_{int}\sim b^{\dagger}|\downarrow\rangle\langle\uparrow|+b|\uparrow\rangle\langle\downarrow|,
\end{equation}
where $b^{(\dagger)}$ are the annihilation (creation) operators for the field modes of the thermal field and the Hamiltonian is only acting on the subspace of the first spinor. A first order interaction will turn the entangled state:
\begin{align}
|\Psi\rangle&=\tfrac{1}{\sqrt{2}}\left(|\uparrow\uparrow\rangle+|\downarrow\downarrow\rangle\right)|n_0\rangle,
\end{align}
into:
\begin{align}
|\Psi\rangle&\rightarrow 
\tfrac{1}{\sqrt{2}}\left(|\downarrow\uparrow\rangle| n_0+1\rangle+|\uparrow\downarrow\rangle | n_0-1\rangle\right).
\end{align}
Since one is only observing the spin degrees of freedom the field subspace must be traced out. One then finds the spins disentangled, i.e.,
\begin{equation}
\text{Tr}_B~\rho(\Psi)\rightarrow\tfrac{1}{2}|\uparrow\downarrow\rangle\langle\uparrow\downarrow|+\tfrac{1}{2}|\downarrow\uparrow\rangle\langle\downarrow\uparrow|.
\end{equation}

In this paper we show that both of these effects can be incorporated into a single framework by using an open quantum system formalism and extending the single electron work of Bell and Leinaas to an entangled two particle system. In our case we take an accelerating electron similar to the one in \cite{Bell:1982qr}, and entangle it with an ancillary rest particle. We will show that the first effect (described above) is nothing other than the Thomas precession of the electron, which is automatically present as a term in the effective Hamiltonian \cite{Bell:1982qr}. In section \ref{sec:evolution} we show that the first effect can be neglected for the classical path under our consideration. The second effect is then the dominant source of disentanglement and within the open system framework we show that analytical values for the relaxation timescales and the concurrence of the system can be obtained. \\

It is found that due to the Unruh radiation that the accelerated particle experiences, the acceleration increases the rate at which the entanglement is destroyed. In the limit when the Rindler temperature (\ref{eqn:rindtemp}) is larger than the spin energy separation scale this dependence takes on a particularly simple form. For an initially maximally entangled system the time to completely disentangle, as measured by the accelerated spinor, is found to be:\\

\bea
\tau_0=\frac{3\pi \ln 3}{8}\frac{\hbar c^6}{\mu^2a^3},
\eea
and in the rest frame:
\beq
t_0=\frac{c}{2a}\text{exp}\left(\frac{3\pi \ln 3}{8}\frac{\hbar c^5}{\mu^2 a^2}\right).
\eeq

This paper is organised as follows: after briefly reviewing accelerated world lines on flat spacetimes in section \ref{sec:review} we outline our configuration of fields and electrons and then determine and solve the master equation for this system in section \ref{sec:evolution}. In section \ref{sec:concurrence} we will calculate the relaxation times and study the time decay of the concurrence in the system before summarising our findings.
\section{Accelerated paths in one dimension \label{sec:review}}
In what follows we give a brief derivation of the Rindler worldline emphasising its generalisation to time-dependent accelerations, $a(\tau)$. It is for this reason we expect that the method we use to calculate concurrence in this paper will also be applicable to more general accelerations, for instance sinusoidal motion.\\ 

Consider a particle confined to move along the $z$-direction and subjected to time-dependent accelerations in its 
IRF
. In that frame the particle will gain a small velocity $dv'=a(\tau)d\tau$ in a small time interval $d\tau$ due to its acceleration, where $a(\tau)$ is the acceleration as measured in the IRF. A static observer will measure a velocity
\beq
v+dv=\frac{dv'+v}{1+v dv'/c^2},
\eeq
where $v$ is the velocity of the IRF at time $t(\tau)$. Keeping terms to order $\mathcal{O}(dv')$ one finds $dv=(1-(v/c)^2)dv'$ and that the rapidity is given by:
\bea
r(\tau)&\equiv&\text{~arctanh}\left(\frac{v(\tau)}{c}\right),\\
&=&\frac{1}{c}\int\frac{dv}{1-(v(\tau)/c)^2}=\frac{1}{c}\int a(\tau)d\tau,
\eea
The particle's trajectory in the static frame is then
\begin{eqnarray}
z(\tau)&=&\int\frac{dz}{dt}\frac{dt}{d\tau}d\tau,\\
&=&c \int \sinh r(\tau) d\tau.\label{eqn:ztau}
\end{eqnarray} 
We can also find the dependence of coordinate time on the proper time, $t(\tau)$, in flat spacetime using the Minkowski line element $1=(\frac{dt}{d\tau})^2-(\frac{dz}{c d\tau})^2$. This implies that
\begin{eqnarray} t(\tau)=\int\cosh r(\tau)d\tau .\label{eqn:ttau}
\end{eqnarray}
For the special case where the acceleration is constant,
$a(\tau)=a$, we obtain:
\begin{eqnarray}\label{eqn:rindlert}
t(\tau)=\frac{c}{a}\sinh\frac{a}{c}\tau, \\ \label{eqn:rindlerz}
 z(\tau)=\frac{c^2}{a}\cosh\frac{a}{c}\tau,
\end{eqnarray}
which is the usual Rindler result. The path of the Rindler worldline is shown in figure \ref{fig:accn_electron}.(a).
\section{Quantum evolution of spin entanglement between frames}\label{sec:evolution}
\begin{figure}[t]
\centering
\fbox{
\includegraphics[scale=.91]{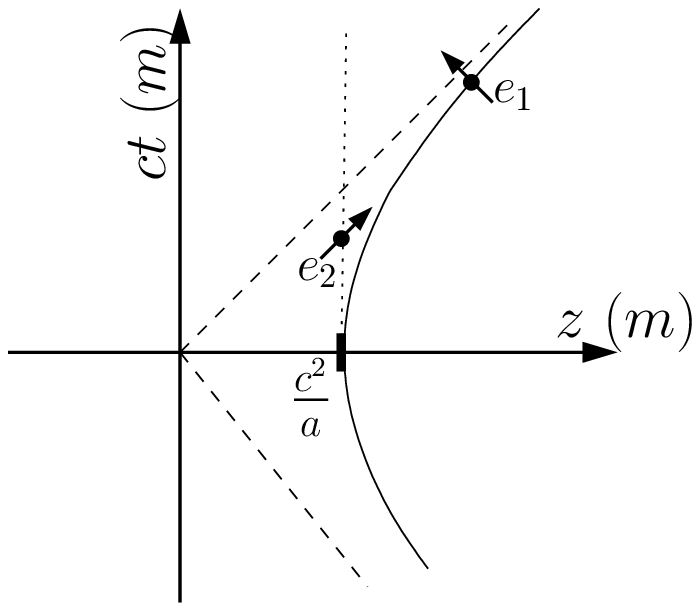}{a)}
}
\fbox{
\includegraphics[scale=0.5]{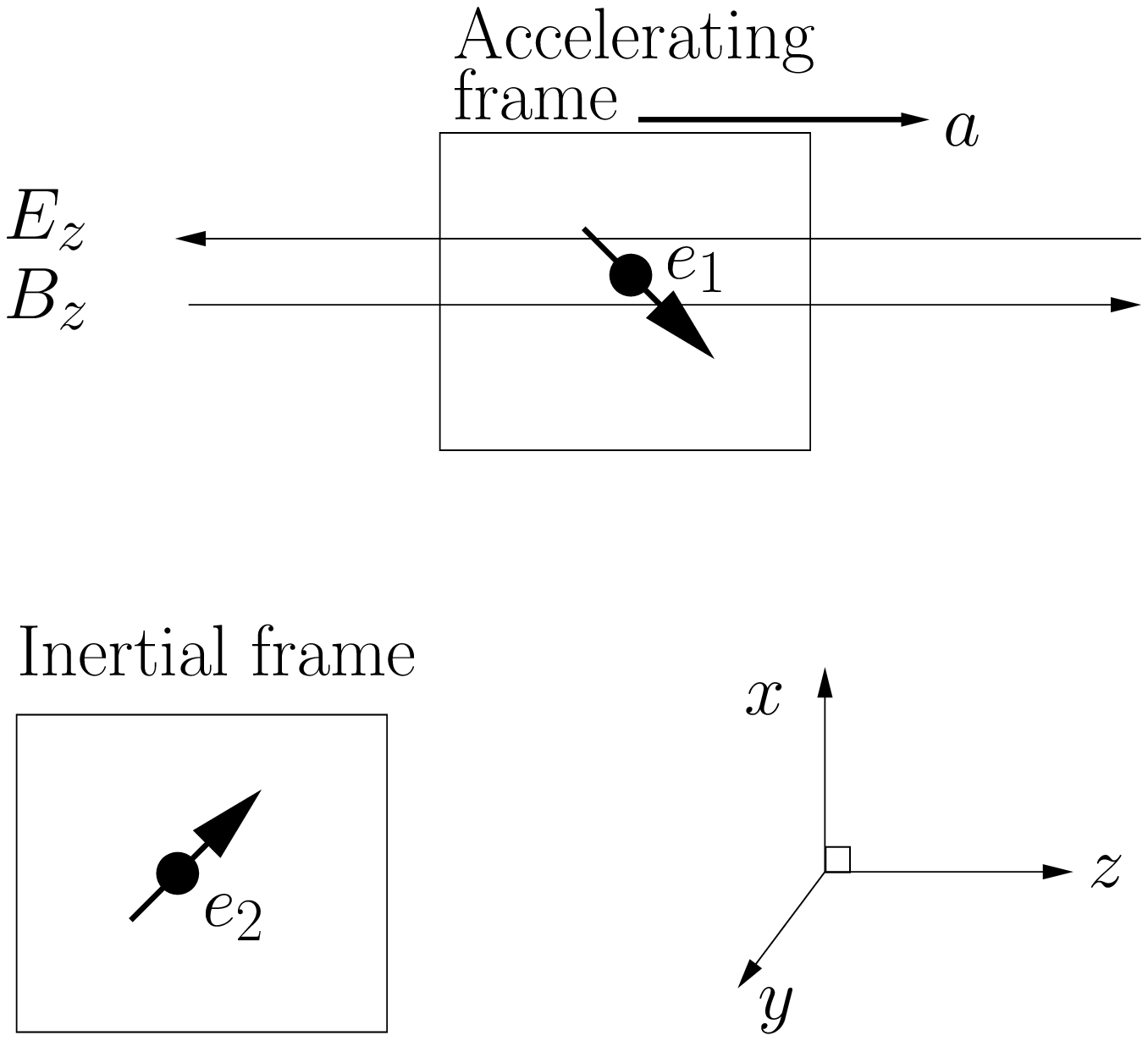}{b)}
}
\caption{Two different views of the two entangled spins in our setup. In a) we have the Minkowski diagram with electron $e_1$ following the Rindler worldline for a constantly accelerated path while electron $e_2$ remains at rest (following the vertical dotted line) and defines the inertial frame of this diagram. Figure b) on the other hand is a physical view of the two electrons;  the top electron, $e_1$, is accelerated by a constant electric field that points in the negative z-direction and placed under a constant magnetic field that points in the positive z-direction, while the bottom spinor, $e_2$, is isolated and at rest.}
\label{fig:accn_electron}
\end{figure}
In our setup, we take two electrons keeping one of them ($e_2$) stationary and isolated from any fields whilst the other ($e_1$) is placed under a constant magnetic field and accelerated by a constant electric field, see figure \ref{fig:accn_electron}. In this situation all the dynamics occurs on $e_1$ with $e_2$ playing a spectator role. $e_2$ acts as an entangling partner for $e_1$ and also defines a static inertial frame. We investigate what effect the acceleration has on any entanglement the two electron spins initially possess.\\

Let the electron mass be $m$, charge be $e$ and the magnitude of the magnetic moment be $\mu$, and define the fields to point along the $z$-direction i.e., $\mathbf{E}=(0,0,-E_z)$ and $\mathbf{B}=(0,0,B_z)$. The IRF acceleration of the first electron $e_1$ is given by:
\begin{equation}
a=-\frac{e}{m} E_z.
\end{equation}
Since the electron charge is negative the electron accelerates in the positive z-direction shown in figure \ref{fig:accn_electron}.(b).\\

The evolution of the electron spin is determined using classical relativistic electrodynamics (see Appendix A and also \cite{Jackson:1975qi,Jackson:1998,Bell:1982qr}). The effective spin-field interaction Hamiltonian for the spin of a relativistic electron \cite{foot:EffHam} is:
\begin{equation} \label{eqn:HRelecKinematic}
H=-\bm{\sigma}\cdot\left\{\gamma^{-1}\mu \bm{B'}-\frac{\hbar}{2}\frac{\gamma^2}{\gamma+1}\frac{d\bm{\beta}}{dt}\times \bm{\beta}\right\},
\end{equation}
where $\bm{\beta}=\bm{v}/{c}$, $\gamma=(1-\beta^2)^{-1/2}$, $\bm{\sigma}$ is the vector of Pauli matrices, and $\bm{B}'$ is the magnetic field in the IRF given by:
\begin{equation}
\bm{B'}=\gamma\left\{\bm{B}-\bm{\beta}\times \bm{E}-\frac{\gamma}{\gamma+1}\bm{\beta}\bm{\beta}\cdot\bm{B}\right\}.
\end{equation}
As discussed in Appendix A the second term in equation (\ref{eqn:HRelecKinematic}) is related to the phenomenon of Thomas precession.\\

We first analyse this equation semi-classically, treating the magnetic field as a classical field and taking the electron path to be the classical trajectory given by equations (\ref{eqn:rindlert}) and (\ref{eqn:rindlerz}). At this level the Thomas term vanishes as the cross product between the velocity of the classical particle and its acceleration is zero. 
Changing the time parameter to the proper time, $\tau$, (corresponding to multiplying the Hamiltonian (\ref{eqn:HRelecKinematic}) by $\gamma$) the evolution of the spin is determined by the usual semi classical Hamiltonian 
\begin{equation}\label{eqn:ClassicalHamiltonian}
H_{\rm SC}=-\mu \sigma_zB_z,
\end{equation}
where we have used the result that $B_z'=B_z$ for the classical path and field. This time-independent system has two energy eigenstates defined by
$\sigma_z=\pm 1$, with an energy gap $\Delta= 2 \mu B_z$ and therefore spin flipping does not occur at this level.\\

From the Lorentz force equation (assuming that all non electromagnetic forces are negligible), one finds:
\begin{eqnarray}
\dot{\bm{p}}&=& mc(\dot{\gamma}\bm{\beta}+\gamma \dot{\bm{\beta}}),\\
 &=& e(\bm{E}+\bm{\beta}\times \bm{B}),
\end{eqnarray}
and thus equation (\ref{eqn:HRelecKinematic}) can be written as:
\begin{equation} \label{eqn:relelecH}
H=-\bm{\sigma}\cdot\left\{\gamma^{-1}\mu \bm{B'}+(\gamma+1)^{-1}\frac{e\hbar}{2mc}\bm{\beta}\times \bm{E'}\right\},
\end{equation}
using:
\begin{equation}
\bm{E'}=\gamma\left(\bm{E}+\bm{\beta}\times\bm{B}-\frac{\gamma}{\gamma+1} \bm{\beta} (\bm{\beta}\cdot \bm{E})\right). 
\end{equation}
In terms of the fields in the stationary frame,
\begin{equation}
H=-\mu \bm{\sigma}\cdot\left\{\gamma^{-1} \bm{B}+(\gamma+1)^{-1}\bm{E}\times\bm{\beta}\right\},
\end{equation}
where $\mu=\frac{e\hbar}{2mc}$ with $g=2$, which differs from (\ref{eqn:relelecH}) by a minus sign in the second term. Using $\bm{p}=mc\gamma \bm{\beta}$ and the classical to quantum correspondence $\hat{\bm{p}}\rightarrow -i\hbar \nabla$ we find:
\begin{equation}\label{eqn:HamiltNabla}
H=-\mu \bm{\sigma}\cdot\left\{\bm{B}-i(\gamma+1)^{-1}\frac{\hbar}{mc}\bm{E}\times\nabla\right\},
\end{equation}
where we have parameterized the quantum evolution by the proper time, $\tau$. Thus, even at the classical field level, if one was to consider a spread in the momentum of the electron \cite{PhysRevLett.89.270402,Peres:2002ip}, the Thomas term would alter the energy separation found in equation (\ref{eqn:ClassicalHamiltonian}). Furthermore, the electric field vacuum fluctuations perpendicular to the momentum would cause the particle to fluctuate about the classical path. Following the discussions relating to the perturbations of the classical path made in reference \cite{Bell:1982qr} we assume, as they have, that the particle path remains classical and unperturbed. This seems to be a good approximation for particles with a gyromagnetic factor $g\sim 2$, such as electrons. Since we neglect perturbations to the path, the Thomas term in equation (\ref{eqn:HamiltNabla}) should be dropped and the moving electron will remain within this approximation in a momentum eigenstate along the z-direction. \\

Thus far we have only been concerned with the dynamics between $e_1$ and the electromagnetic fields, we now introduce the ancillary electron $e_2$ into the Hamiltonian. We take the Hilbert space of the combined system to be $\mathcal{H}=\mathcal{H}_1\otimes\mathcal{H}_2\otimes\mathcal{H}_{\text{field}}$,
where $\mathcal{H}_1$ is the subspace of the accelerated
spinor, $\mathcal{H}_2$ is the subspace of the spectator electron, and $\mathcal{H}_{\text{field}}$ is the Hilbert space for the electromagnetic field. In terms of the proper time the spin-field interaction Hamiltonian for the whole system is:
\begin{equation}\label{eqn:reducedHamiltonian}
H=-\mu \bm{\sigma}\otimes\mathbbm{1}\cdot\bm{B},
\end{equation}
Since the action on $\mathcal{H}_2$ is always the identity, for notational simplicity we suppress the factor of $\otimes \mathbbm{1}$ from all operators acting on the combined spin Hilbert space in what follows.\\

We now consider the second quantisation of the magnetic field. We do this in the standard way by interpreting the magnetic field as a space-time dependent field operator. Expanding equation (\ref{eqn:reducedHamiltonian}), and defining the field operators, $B_{\pm}=B_{x}\pm i B_{y}$, \cite{basischange} we obtain the interaction in terms of field mode operators:
\begin{eqnarray}
V(x)&=&-\mu (\sigma_x B_x(x)+\sigma_y B_y(x)+\sigma_z B_z(x)),\nonumber\\
&=&-\mu(\sigma_-B_+(x)+\sigma_+B_-(x)+\sigma_z B_z(x)),\label{eqn:qfspin-field}
\end{eqnarray}
where in the last step we have used $\sigma_{\pm}=\frac{1}{2}(\sigma_x\mp i\sigma_y)$.
Thus, the spin flips when a magnetic field mode \cite{photonrel} is excited.\\

Next we calculate the time evolution of the two electron system using a perturbative master equation \cite{paz-2000, 535Walls} whereby we interpret the magnetic field fluctuations about this system as an external environment \cite{justifyperturb}. As we are viewing the fluctuations of the external magnetic field as an unobserved environment of the spin-spin system, tracing over the field subspace is implied and thus no \textit{ad hoc} dynamical modelling of the source of decoherence is necessary.\\

We can write the total Hamiltonian of the system as
\beq
H=H_{SC}+H_{B}+V,
\eeq
where $H_{SC}$ is the semiclassical evolution of the spin system, (\ref{eqn:ClassicalHamiltonian}), $H_B$ is the free Hamiltonian for the magnetic field, whose exact form will not be required, and V contains the quantum fluctuations arising from the second quantization of the spin-field interaction determined in equation (\ref{eqn:qfspin-field}).\\

Let $\rho_T$ be the total density operator of the system plus field in the interaction picture. The equation of motion is:
\beq\label{eqn:totaldensitySE}
\frac{d\rho_T(\tau)}{d\tau}=\frac{1}{i\hbar}[V(\tau),\rho_T(\tau)],
\eeq
where $\tau$ is the proper time of the accelerating electron. We assume that initially the electron spin system and field states are uncorrelated so that:
\beq
\rho_T(0)=\rho_I(0)\otimes\rho_B,
\eeq
where $\rho_I$ describes the state of the two electron spins and $\rho_B$ describes the state of the magnetic field. Following \cite{535Walls}, one can expand equation (\ref{eqn:totaldensitySE}) in a perturbative series and noting that $\text{Tr}_B(V(\tau)\rho_B)=0$ \cite{tracezero} one finds to second order in perturbation theory:
\beq\label{eqn:2ndOrderMaster}
\frac{d\rho_I(\tau)}{d\tau}=-\frac{1}{\hbar^2}\int_0^\tau d\tau_1\text{Tr}_B[V(\tau),[V(\tau_1),\rho_I(\tau)\otimes \rho_B]].
\eeq
For brevity we write the interaction in (\ref{eqn:qfspin-field}) as
\beq
V(\tau)=-\mu \sum_i\sigma_i(\tau) B^{\dagger}_i(x(\tau)),
\eeq
where $i=\{+,-,z\}$ and the $\sigma_i(\tau)$ are Heisenberg operators. By taking out the dependence of $H_{SC}$ from $\rho_I$ using:
\begin{equation}
\rho_I(\tau)=e^{iH_{SC}\tau / \hbar}\rho_S(\tau) e^{-i H_{SC}\tau/\hbar},
\end{equation}
we can rewrite the master equation (\ref{eqn:2ndOrderMaster}) for the spin subsystem in terms of $\rho_S$, i.e., the reduced density matrix in the Schr\"odinger picture (see \cite{paz-2000}):
\begin{align}
\frac{d\rho_S(\tau)}{d\tau}&=\frac{1}{i\hbar}[H_{SC},\rho_S]-\frac{\mu^2}{\hbar^2}\sum_{i,j}\int_0^\tau d\tau_1\times\cdots\nonumber\\&\text{Tr}_B[\sigma_i B^{\dagger}_i(x),[\sigma_j(\tau_1-\tau)B_j^{\dagger}(x_1),\rho_S\otimes \rho_B]],
\end{align}
where $x\equiv x(\tau)$ and $x_1 \equiv x(\tau_1)$ are defined for short. This equation is still not entirely in the spin subsystem Schr\"odinger picture as we still have the $\sigma(\tau_1-\tau)$ Heisenberg operators. However since $H_{SC}$ is given by equation (\ref{eqn:ClassicalHamiltonian}) we can use the Heisenberg equations of motion to find
\begin{equation}
\sigma_{j}(\tau)=e^{\alpha_j \Delta i \tau/\hbar}\sigma_j,
\end{equation}
where $\alpha_j={+1,-1,0}$, for $j=+,-,z$, respectively, and defining the expectation value over the field state to be $\langle O\rangle=\text{Tr}_B(O\rho_B)$ we obtain:
\begin{align}
\frac{d\rho_S}{d\tau}&=\frac{1}{i\hbar}[H_{SC},\rho_S]-\frac{\mu^2}{\hbar^2}\sum_{i,j}\int_0^\tau d\tau_1e^{\alpha_j \Delta i (\tau_1-\tau)/\hbar}\times\nonumber\\&\left\{(\sigma_i\sigma_j\rho_S-\sigma_j\rho_S\sigma_i)
\langle B_i^{\dagger}(x)B_j^{\dagger}(x_1)\rangle+\right.\nonumber\\
&~~~~~~+\left.(\rho_S\sigma_j\sigma_i-\sigma_i\rho_S\sigma_j)
\langle B_j^{\dagger}(x_1)B_i^{\dagger}(x)\rangle \right\}.
\end{align}
In Gaussian units the electromagnetic free
field Wightman function is
\begin{align}
\langle 0 |&F_{\mu\nu}(x)F_{\rho\sigma}(x')|0 \rangle=\frac{4\hbar
c}{\pi}(x-x')^{-6}\times\nonumber\\
&\left\{(x-x')^2(g_{\mu \rho}
g_{\nu\sigma}-g_{\mu\sigma}g_{\nu\rho})\right.\nonumber
\\&\left.-2\left[(x-x')_\mu(x-x')_\rho g_{\nu\sigma}-(x-x')_\nu(x-x')_\rho
g_{\mu\sigma}\right.\right.\nonumber\\
&\left.\left.-(x-x')_\mu(x-x')_\sigma
g_{\nu\rho}+(x-x')_\nu(x-x')_\sigma g_{\mu\rho}\right]\right\},
\end{align}
where $F_{\mu\nu}=\partial_{\mu}A_{\nu}-\partial_{\nu}A_{\mu}$ is the anti-symmetric electromagnetic field strength tensor. Using $A^{\mu}=(\phi,\vec{A})$ and $\vec{B}=\nabla \times \vec{A}$ one finds $\vec{B}=(F_{32},F_{13},F_{21})$. Since the motion is entirely along the $z$ direction we find that 
the only non-zero two point correlation functions are \cite{epsilon}
\beq
\langle B_j(x)B_j(x')\rangle=G(x(\tau-i\epsilon),x(\tau ')),
\eeq
where $j\in\{{x,y,z}\}$ \cite{othercorrelations} and 
\begin{equation}
G(x,x')= \frac{4 \hbar c}{\pi} (x-x')^{-4}.
\end{equation}
Along the constantly accelerated path described by equations
(\ref{eqn:rindlert})-(\ref{eqn:rindlerz}), this further reduces to:
\begin{eqnarray}\label{eqn:G}
G(\tau-\tau ')&\equiv&G(x(\tau),x(\tau '))\nonumber\\
&=&\frac{\hbar a^4}{4\pi
c^7}\left\{\sinh\left[\frac{a}{2c}(\tau-\tau')\right]\right\}^{-4}.
\end{eqnarray}
Thus we obtain the equation:
\begin{widetext}
\begin{align}\label{eqn:masterlong}
\frac{d\rho_S}{d\tau}&=\frac{1}{i\hbar}[H_{SC},\rho_S]-\frac{\mu^2}{\hbar^2}\left\{\begin{array}{c}~\\~\end{array}~~\right.\nonumber\\
&2(\sigma_-\sigma_+\rho_S-\sigma_+\rho_S\sigma_-)\int_0^\tau d\tau_1 e^{i\Delta (\tau_1-\tau)/\hbar} G(\tau-\tau_1-i\epsilon)
+2(\rho_S\sigma_+\sigma_--\sigma_-\rho_S\sigma_+)\int_0^\tau d\tau_1 e^{i\Delta (\tau_1-\tau)/\hbar}G(\tau-\tau _1+i\epsilon)\nonumber\\
&+2(\sigma_+\sigma_-\rho_S-\sigma_-\rho_S\sigma_+)\int_0^\tau d\tau_1 e^{-i\Delta (\tau_1-\tau)/\hbar} G(\tau-\tau_1-i\epsilon)
+2(\rho_S\sigma_-\sigma_+-\sigma_+\rho_S\sigma_-)\int_0^\tau d\tau_1 e^{-i\Delta (\tau_1-\tau)/\hbar}G(\tau-\tau _1+i\epsilon)\nonumber\\
&\left.\begin{array}{c}~\\~\end{array}+(\rho_S-\sigma_z\rho_S\sigma_z)\int_0^\tau\left(G(\tau-\tau_1-i\epsilon)+G(\tau-\tau_1+i \epsilon)\right)d\tau_1\right\}.
\end{align}
\end{widetext}
Since the integrands in the above equation are sharply peaked functions about $\tau_1=\tau$ we are justified in making the Markovian approximation (see \cite{paz-2000}, pg.28). We make the change of variable $s=\tau-\tau_1$ and extend the integration over the interval $s\in [0,\infty]$. To simplify the notation we define the following integrals:
\bea
\Gamma_{\pm}&=&\int_0^{\infty}e^{i\Delta s/\hbar}G(s\pm i\epsilon)ds,\\
\Gamma_z&=&\int_0^{\infty}\left\{G(s-i\epsilon)+G(s+i \epsilon)\right\}ds.
\eea
Using
\beq
\int \frac{ds}{\sinh^4(s+i a)}=-\tfrac{1}{3}\coth(s+i a)\left(\text{csch}^2(s+i a)-2\right),
\eeq
we find that $\Gamma_z=\frac{2}{3}\frac{\hbar a^3}{\pi c^6}$.
Then equation (\ref{eqn:masterlong}) becomes:
\begin{align}\label{eqn:mastermed}
\frac{d\rho_S}{d\tau}&=\frac{1}{i\hbar}[H_{SC},\rho_S]-\frac{\mu^2}{\hbar^2}\left\{\right.\nonumber\\
&2(\sigma_-\sigma_+\rho_S-\sigma_+\rho_S\sigma_-)\Gamma_+^*
+2(\rho_S\sigma_+\sigma_--\sigma_-\rho_S\sigma_+)\Gamma_-^* \nonumber\\
+&2(\sigma_+\sigma_-\rho_S-\sigma_-\rho_S\sigma_+)\Gamma_-
+2(\rho_S\sigma_-\sigma_+-\sigma_+\rho_S\sigma_-)\Gamma_+\nonumber\\
+&\frac{2}{3}\frac{\hbar a^3}{\pi c^6}(\rho_S-\sigma_z\rho_S\sigma_z)\left.\right\},
\end{align}
where we have made use of the identity $G(-z)=G(z)$ for any complex $z$, which follows from equation (\ref{eqn:G}). This master equation is similar in form to those found in \cite{paz-2000, Fisher}.  By separating $\Gamma_1$ and $\Gamma_2$ into their real and imaginary components equation (\ref{eqn:mastermed}) simplifies into:
\begin{align}\label{eqn:mastersmall}
\frac{d\rho_S}{d\tau}&=\frac{1}{i\hbar}[H_{SC},\rho_S]-\frac{\mu^2}{\hbar^2}\left\{\right.\nonumber\\
&2\text{Re} \Gamma_- (\sigma_-\sigma_+\rho_S+\rho_S\sigma_-\sigma_+-2\sigma_+\rho_S\sigma_-)
\nonumber\\
+&2\text{Re}\Gamma_+(\sigma_+\sigma_-\rho_S+\rho_S\sigma_+\sigma_--2\sigma_-\rho_S\sigma_+)\nonumber\\
+&\frac{2}{3}\frac{\hbar a^3}{\pi c^6}(\rho_S-\sigma_z\rho_S\sigma_z)-i\text{Im}(\Gamma_++\Gamma_-)[\sigma_z,\rho_S]\left.\right\}.
\end{align}
The last term is a correction to the unperturbed energy separation $\Delta$. It is due to the spin-field coupling and effectively renormalizes the Hamiltonian by a term $H_{LC}=\frac{\mu^2}{\hbar}\text{Im}(\Gamma_++\Gamma_-)\sigma_z$, reminiscent of the Lamb shift \cite{SakauriAdv} in atomic physics.  Further defining:
\bea\label{eqn:fliprates}
\gamma_\pm \equiv\frac{4\mu^2}{\hbar^2}\text{Re}\Gamma_\pm=\frac{2\mu^2}{\hbar^2}\int_{-\infty}^{\infty} e^{\mp i\Delta s/\hbar} G(s-i\epsilon)ds,
\eea
we observe that the Master equation takes a manifestly Lindblad form: 
\begin{equation}
\frac{d\rho_S}{dt}=-\frac{i}{\hbar} [H',\rho_S]+\sum_j\left[2 L_j\rho_S
L_j^\dagger-\left\{L_j^\dagger L_j,\rho_S\right\}\right],
\end{equation}
where $\{x,y\}=xy+yx$ denotes an anticommutator, $H'=H_{SC}-H_{LC}$ is the coherent part of the renormalized spin Hamiltonian, and $L_j$ are the Lindblad operators, given by:
\bea
L_1 &=&\sqrt{\frac{\gamma_-}{2}} \sigma_-,\\ 
L_2 &=&\sqrt{\frac{\gamma_+}{2}} \sigma_+,
\eea
for transitions down and up (in spin energy) respectively and 
\beq
L_3=\sqrt{\frac{\gamma_z}{2}}\sigma_z,
\eeq
(a pure dephasing channel) where 
\beq\label{eqn:gammaz}
\gamma_z\equiv \frac{\mu^2}{\hbar^2}\Gamma_z= \frac{2}{3}\frac{\mu^2}{\hbar} \frac{a^3}{\pi c^6}.
\eeq
By renormalizing the subsystem density matrix,
\begin{equation}\label{eqn:rhoInt}
\tilde{\rho}(\tau)=e^{iH' \tau/\hbar}\rho_S(\tau)e^{-iH' \tau/\hbar},
\end{equation}
we can write
\begin{eqnarray}\label{eqn:Masterunfinished}
\frac{d\tilde{\rho}}{d\tau}&=&\tfrac{\gamma_-}{2} \left[ 2
\sigma_-\tilde{\rho}\sigma_+-\sigma_+\sigma_-\tilde{\rho}-\tilde{\rho}\sigma_+\sigma_-\right]\nonumber\\
&+&\tfrac{\gamma_+}{2} \left[ 2
\sigma_+\tilde{\rho}\sigma_--\sigma_-\sigma_+\tilde{\rho}-\tilde{\rho}\sigma_-\sigma_+\right]\nonumber\\
&+&\gamma_z[\sigma_z\tilde{\rho}\sigma_z-\tilde{\rho}].
\end{eqnarray}
We note that this master equation is similar to the master equation for spontaneous emission of an atom discussed in \cite{IkeAndMike} (pg. 388) except that in our case we have a $4\times 4$ density matrix and two more Lindblad operators. Nevertheless, we will now show that the same method that was used to solve the master equation in \cite{IkeAndMike} can be applied here by choosing the right generalisation of the Bloch sphere in $4\times 4$ dimensions.\\

Since $\rho$ spans a sixteen dimensional vector space and the direct product of Pauli matrices including the identity, $\{\sigma_i\otimes\sigma_j|{i,j}\in{0,\cdots,3}\}$, form sixteen linearly independent vectors we can expand any general density matrix for a two spin system as follows:
\begin{equation}\label{eqn:Arbintialdensity}
\rho=\sum_{i=0}^3\sum_{j=0}^3r_{ij}\sigma_i\otimes\sigma_j,
\end{equation}
where we have chosen the Pauli matrices:
\begin{equation}
\sigma_{1}=\left(
\begin{array}{cc}
0 & 1 \\ 1 & 0
\end{array}
\right) ,\ \ \ \sigma_{2}=\left(
\begin{array}{cc}
0 & -i \\ i & 0
\end{array}
\right),\ \ \ \sigma_{3}=\left(
\begin{array}{cc}
1 & 0 \\ 0 & -1
\end{array}
\right) ,
\end{equation}
and defined $\sigma_0=\mathbbm{1}$. A nice property about this choice of basis is that the expansion coefficients $r_{ij}$ are real, which follows from the hermiticity of the Pauli matrices and density operator, furthermore the expansion coefficients can be computed directly using:
\begin{equation}
r_{ij}=\frac{1}{4}\text{Tr}(\rho \sigma_i\otimes \sigma_j).
\end{equation}
As every density matrix has trace one, $r_{00}$ is equal to one quarter. From the inequality $\text{Tr} \rho^2\leq 1$ the density operator can also be expressed as:
\begin{equation} \label{eqn:generalisedBloch}
\rho=\frac{1}{4}\left(\mathbf{1}_{4\times 4 }+\sqrt{3}\sum_{i+j\neq 0}y_{ij}\sigma_i\otimes\sigma_j\right),
\end{equation}
where 
\begin{equation}
\sum_{i+j\neq 0}(y_{ij})^2\leq 1,
\end{equation}
where equality holds if and only if the state is pure. Equation (\ref{eqn:generalisedBloch}) generalises the Bloch sphere representation of a single qubit \cite{IkeAndMike} to a two qubit system, in this way two qubit mixed states can be thought of as lying somewhere within a fifteen dimensional unit-sphere.\\

In our case we will find it more useful to express our density in the form (\ref{eqn:Arbintialdensity}). As an example, a maximally entangled Bell state of the kind $\tfrac{1}{\sqrt{2}}\left(|\uparrow\uparrow\rangle+|\downarrow\downarrow\rangle\right)$ would be expressed
\beq\label{eqn:rhobell}
\rho_{\rm Bell}=\tfrac{1}{4}(\sigma_0\otimes\sigma_0+\sigma_1\otimes\sigma_1-\sigma_2\otimes\sigma_2+\sigma_3\otimes\sigma_3).
\eeq

The entanglement between the spin degrees of freedom can be obtained by calculating the concurrence \cite{Wootters:1997id}
\beq \label{eqn:con1}
C(\rho)=\text{max}\{\lambda_1-\lambda_2-\lambda_3-\lambda_4,0\},
\eeq
where $\{\lambda_1,\lambda_2,\lambda_3,\lambda_4\}$ are the non-negative square roots of the eigenvalues of the matrix
\beq\label{eqn:con2}
M=\rho (\sigma_2\otimes\sigma_2)\rho^*(\sigma_2\otimes\sigma_2).
\eeq
Thus, the maximally entangled state (\ref{eqn:rhobell}) has a concurrence of one.\\

Substituting $\tilde{\rho}=\sum_{i,j}\tilde{r}_{ij}\sigma_{i}\otimes\sigma_{j}$ into equation (\ref{eqn:Masterunfinished}) gives:
\begin{align}
\frac{d\tilde{r}_{ij}}{d\tau}\sigma_{i}\otimes\sigma_{j}&=\frac{\gamma_-}{2} \tilde{r}_{ij}\left[ 2
\sigma_-\sigma_{i}\sigma_+-\sigma_+\sigma_-\sigma_{i}-\sigma_{i}\sigma_+\sigma_-\right]\otimes \sigma_{j}\nonumber\\&+\frac{\gamma_+}{2}\tilde{r}_{ij}\left[ 2
\sigma_+\sigma_{i}\sigma_--\sigma_-\sigma_+\sigma_{i}-\sigma_{i}\sigma_-\sigma_+\right]\otimes\sigma_{j}\nonumber\\
&+\gamma_z \tilde{r}_{ij}[\sigma_z\sigma_{i}\sigma_z-\sigma_{i}]\otimes\sigma_j,
\end{align}
which after a little algebra gives sixteen first order linear differential equations
\bea
\dot{\tilde{r}}_{0j}(\tau)&=&0,\nonumber\\
\dot{\tilde{r}}_{1j}(\tau)&=&-\tfrac{1}{2}(\gamma_-+\gamma_++4\gamma_z)\tilde{r}_{1j}(\tau),\nonumber\\
\dot{\tilde{r}}_{2j}(\tau)&=&-\tfrac{1}{2}(\gamma_-+\gamma_++4\gamma_z)\tilde{r}_{2j}(\tau),\nonumber\\
\dot{\tilde{r}}_{3j}(\tau)&=&(\gamma_--\gamma_+)\tilde{r}_{0j}(\tau)-(\gamma_-+\gamma_+)\tilde{r}_{3j}(\tau),\label{eqn:rdot1}
\eea
where dots imply differentiation with respect to $\tau$.
The solutions to these equations are found to be:
\begin{align}\label{eqn:solution}
\tilde{r}_{0j}(\tau)&=\tilde{r}_{0j}(0),\nonumber \\
\tilde{r}_{1j}(\tau)&=\tilde{r}_{1j}(0)e^{-\tfrac{1}{2}(\gamma_- +\gamma_++4\gamma_z)\tau},\nonumber \\
\tilde{r}_{2j}(\tau)&=\tilde{r}_{2j}(0)e^{-\tfrac{1}{2}(\gamma_- +\gamma_++4\gamma_z)\tau}, \nonumber\\
\tilde{r}_{3j}(\tau)&=\tilde{r}_{3j}(0)e^{-(\gamma_-+\gamma_+)\tau}\nonumber\\
&~~~+\tfrac{\gamma_--\gamma_+}{\gamma_-+\gamma_+}\tilde{r}_{0j}(0)(1-e^{-(\gamma_-+\gamma_+)\tau}).
\end{align}
Thus, the relaxation and dephasing times respectively are:
\bea\label{eqn:T1inverse}
T_1^{-1}&=&\gamma_-+\gamma_+,\\
T_2^{-1}&=&\frac{\gamma_-+\gamma_+}{2}+2\gamma_z,\label{eqn:T2inverse}
\eea
where $T_2$ is the timescale for which the off-diagonal elements of the density matrix (`coherences') decay and $T_1$ is the timescale for spin flipping. The relations (\ref{eqn:T1inverse})-(\ref{eqn:T2inverse}) have a form identical to those typically found in magnetic resonance problems (see \cite{Fisher}, chapter 4.3). We will compare these timescales in more detail in the next section once we have calculated the spin flip rates.

\section{Relaxation and Concurrence}\label{sec:concurrence}
We have succeeded in finding the time evolution for the system of spins within our specific setup. In order to analyse the effect that the acceleration has on the entanglement shared by these spins it is necessary to calculate the spin flip rates defined in equation (\ref{eqn:fliprates}).\\

\begin{figure}[t]
\centering
\includegraphics[scale=0.75]{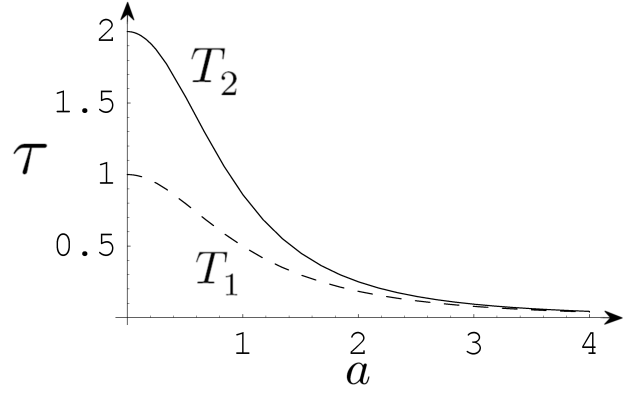}\vspace{-.5cm}\hspace{2cm}
\caption{The relaxation times $T_1$ (dashed line) and $T_2$ (solid line) in units of the spontaneous emission timescale at zero acceleration, $\gamma_0^{-1}$, as a function of the acceleration, $a$, in units of $c\Delta/\hbar$.}
\label{fig:Relax}
\end{figure}

Making a linear change of variables, $s'=as/c$, the integral in
(\ref{eqn:fliprates}) becomes
\begin{equation}\label{eqn:sintegral}
\gamma_{\pm}=\frac{2\mu^2}{\hbar^2}\frac{\hbar
a^4}{4\pi c^7}\frac{c}{a}\int_{-\infty}^{\infty}ds'
\frac{\exp\left( \mp i\Delta s' c/a
\hbar\right)}{\sinh^{4}\tfrac{1}{2}(s'-i\epsilon)}.
\end{equation}
These integrals can be evaluated using contour integration \cite{Bell:1982qr}:
%
\begin{equation}\label{eqn:transitions}
\gamma_{\pm}=\mp\frac{8}{3}
\frac{\mu^2\Delta}{\hbar^4c^3}\left(\Delta^2+\frac{a^2\hbar^2}{c^2}\right)
\left(1-e^{{\pm 2\pi c\Delta/a\hbar}}\right)^{-1}.
\end{equation}
Bell and Leinaas also noted that the ratio of the transition rates,
\begin{equation}\label{eqn:gammaongamma}
\frac{\gamma_+}{\gamma_-}=e^{-2\pi c\Delta /a\hbar},
\end{equation}
define an equilibrium ratio of populations of the upper and lower states. Thus, the equilibrium distribution over the levels has a thermal character in accordance with the Unruh temperature formula (\ref{eqn:rindtemp}). To proceed further we recall that the Bose occupation number is
\beq
n=\frac{1}{e^{2\pi c \Delta/a\hbar}-1}.
\eeq
We also notice from equation (\ref{eqn:transitions}) that there remains an emission rate even when the acceleration goes to zero: 
\beq
\gamma_0\equiv \lim_{a\rightarrow 0}\gamma_- =\frac{8}{3}\frac{\mu^2\Delta^3}{\hbar^4 c^3}.
\eeq
This \textit{spontaneous emission} is due to the quantum interactions of the magnetic moment with the magnetic field and defines a useful parameter, $\gamma_0$, which sets the scale of spin flip transition rates. This magnetic spontaneous emission rate is much weaker than the usual atomic spontaneous emission rate, and is significantly harder to detect, see \cite{gover:124801}. The equations in (\ref{eqn:fliprates}) can now be written:
\bea
\gamma_+&=&\gamma_0\left(1+\left(\tfrac{a\hbar}{c\Delta}\right)^2\right) n,\\
\gamma_-&=&\gamma_0\left(1+\left(\tfrac{a\hbar}{c\Delta}\right)^2\right) (n+1).
\eea
Furthermore, it is easily verified that 
\beq
\gamma_z=\tfrac{1}{4\pi}\gamma_0\left(\tfrac{a\hbar}{c\Delta}\right)^3.
\eeq
Thus, the relaxation and dephasing times respectively are:
\bea
T_1^{-1}&=&\gamma_0(1+(\tfrac{a\hbar}{c\Delta})^2)\coth\tfrac{\pi c\Delta}{a\hbar},\\
T_2^{-1}&=&\tfrac{\gamma_0}{2}\left\{(1+(\tfrac{a\hbar}{c\Delta})^2)\coth\tfrac{\pi c\Delta}{a\hbar}+\tfrac{1}{\pi}(\tfrac{a\hbar}{c\Delta})^3\right\}.\label{eqn:T2}
\eea
The relaxation and dephasing times are shown in figure \ref{fig:Relax}. We observe that $T_1< T_2\leq 2T_1$ for all values of the acceleration. Thus, typically there will be more than one spin flip before the coherences vanish. \\

In the long time limit the state (\ref{eqn:Arbintialdensity}) becomes:
\bea
\tilde{\rho}_a(\infty)&=&\sum_j \tilde{r}_{0j}\mathbf{1}\otimes\sigma_{j}
+\tanh (\tfrac{c\pi\Delta}{a\hbar}) \tilde{r}_{0j}\sigma_{3}\otimes\sigma_{j},\\
&=&\frac{\left(\mathbf{1}+\tanh (\tfrac{c\pi\Delta}{a\hbar})~\sigma_3\right)}{2}\otimes 2 \sum_j \tilde{r}_{0j}\sigma_j,
\eea
which is a product of two density matrices acting separately on each of the spinor subspaces. Therefore the state of the spins will eventually separate regardless of how they were initialised. To determine precisely how the system separates we calculate the concurrence (\ref{eqn:con1})-(\ref{eqn:con2}) as a function of time. Before doing so, it is worth discussing the effect of Lorentz transformations on entanglement.\\

Recall that Gingrich and Adami \cite{PhysRevLett.89.270402} have shown that the concurrence of two particles, each in a state with some spread in momentum, is not Lorentz invariant. This is because momentum dependent Wigner rotations act on the spinors under Lorentz transformations (see also \cite{Peres:2002ip} for a discussion on the entropy of single particles under Lorentz transformations). One then wonders if in the present case calculating the concurrence in the rest frame of the inertial electron is meaningful. While this is clearly a concern for particles with a spread of momentum states, when the particles are in momentum eigenstates the Wigner rotation acts like a local unitary operation on the spinors and therefore does not change the concurrence \cite{PhysRevLett.89.270402, Milburn}. In our case, the moving electron follows the classical path defined by equations (\ref{eqn:rindlert})-(\ref{eqn:rindlerz}) and is thus by construction in a momentum eigenstate $p_z(\tau)$. Furthermore, the stationary spinor can be chosen to be as narrow as is required i.e., $\Delta p_z\sim 0$. Under these conditions, since there is no significant momentum spread in either of the particles wavefunctions, the concurrence function behaves to good approximation invariantly under Lorentz transformations. We now proceed to calculate the value of the concurrence in the stationary electron rest frame.\\

We initialise the system into the maximally entangled Bell state ({\ref{eqn:rhobell}) which then evolves according to equation (\ref{eqn:solution}), since concurrence is invariant under local unitary transformations \cite{Wootters:1997id} we are justified in calculating the concurrence of $\tilde{\rho}$ instead of $\rho_S$. The time-dependent system density matrix is:
\begin{align}\label{eqn:evolution}
\tilde{\rho}(\tau)&=\frac{1}{4}\left\{\sigma_0\otimes\sigma_0+e^{-\Gamma_2 \tau}\sigma_1\otimes\sigma_1-e^{-\Gamma_2 \tau}\sigma_2\otimes\sigma_2\right.\nonumber\\
&\left.+e^{-\Gamma_1 \tau}\sigma_3\otimes\sigma_3+\tanh\left(\tfrac{c\pi\Delta}{a\hbar}\right)\left(1-e^{-\Gamma_1 \tau}\right)\sigma_3\otimes\sigma_0\right\},
\end{align}
where we have defined $\Gamma_1=T_1^{-1}$ and $\Gamma_2=T_2^{-1}$. One observes from equation (\ref{eqn:evolution}) that $\tilde{\rho}(\tau)=\tilde{\rho}^*(\tau)$, and since the eigenvalues of $M$ in equation (\ref{eqn:con2}) are real and positive \cite{Wootters:1997id} we can find the concurrence of $\tilde{\rho}(\tau)$ by diagonalising the matrix:
\beq\label{eqn:newmatrix}
\tilde{\rho}\sigma_2\otimes\sigma_2,
\eeq
where we take the $\{\lambda_i\}$ in equation (\ref{eqn:con1}) to be the absolute values of the eigenvalues of the matrix (\ref{eqn:newmatrix}). We find that the concurrence is given by:
\beq \label{eqn:con3}
C(a,\tau)=\text{max}\left\{e^{-\tau\Gamma_2}-\tfrac{1}{2}\left(1-e^{-\tau\Gamma_1}\right)\text{sech}(\tfrac{c\pi\Delta}{a\hbar}),0\right\}.
\eeq
The concurrence is plotted in figure \ref{fig:concurrence} as a function of the acceleration and the proper time. We observe that the greater the acceleration the quicker that the initial entanglement disappears.
\begin{figure}[h]
\centering
\includegraphics[scale=.62]{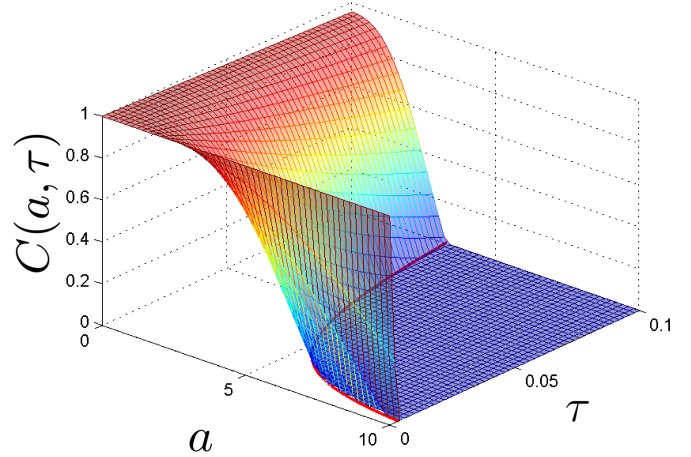}
\caption{(Color online) We have plotted the concurrence of two initially maximally entangled spins as a function of the acceleration, $a$, (in units of $\tfrac{c\Delta}{\hbar}$) and the proper time (in units of $\gamma_0^{-1}$). Our main result, equation (\ref{eqn:main}), is also shown overlaid in a dark (red) line along $C=0$.}
\label{fig:concurrence}
\end{figure}
We will use the time taken for the system to reach zero concurrence to quantify the dependence of disentanglement on acceleration. For a given acceleration the time taken, $\tau_0$, for the system to completely disentangle is given by the equation: 
\beq \label{eqn:taunought}
e^{-\tau_0\Gamma_2}=\tfrac{1}{2}\left(1-e^{-\tau_0\Gamma_1}\right)\text{sech}(\tfrac{c\pi\Delta}{a\hbar}),
\eeq
which follows from equation (\ref{eqn:con3}) \cite{concurrenceclause}.
Defining the dimensionless parameter $\alpha\equiv\frac{a\hbar}{c\Delta}$ and taking the limit $\alpha\gg 1$ (i.e., $\Delta\ll \frac{a\hbar}{c}$) we find, 
\bea
\Gamma_1=\Gamma_2=\frac{8}{3\pi}\frac{\mu^2a^3}{\hbar c^6}+\mathcal{O}(\alpha).
\eea
Equation (\ref{eqn:taunought}) can now be solved for $\tau_0$:
\bea\label{eqn:main}
\tau_0=\frac{3\pi \ln 3}{8}\frac{\hbar c^6}{\mu^2a^3}+\mathcal{O}(\alpha^{-5}).
\eea
Thus in the large acceleration small magnetic field limit the proper time taken to disentangle the two spinors is proportional to the inverse of the acceleration cubed \cite{zeroctime}. We have used this result to plot the zero concurrence dark (red) line in figure \ref{fig:concurrence}.\\
In the frame of the stationary spinor the time taken to disentangle is exponentially longer:
\beq\label{eqn:t0}
t_0=\frac{c}{2a}\text{exp}\left(\frac{3\pi \ln 3}{8}\frac{\hbar c^5}{\mu^2 a^2}\right).
\eeq
If we put in the numbers for an electron we find:
\beq
t_0=\frac{c}{2a}\text{exp}\left(\frac{3.8\times 10 ^{61}\text{m}^2\text{.s}^{-4}}{a^2}\right).
\eeq
Therefore an acceleration with a magnitude of about thirty is required to observe the disentanglement on a reasonable timescale. This is consistent with the thermal equilibrium timescale found in \cite{Bell:1982qr}.
\section{Conclusion}\label{sec:conclusion}
We have considered the system of two spin-entangled electrons when one electron is accelerated and placed under a constant magnetic field whilst the other is at rest and isolated. Our method consisted of explicitly calculating the open quantum system where the quantised magnetic field fluctuations were considered to be an unobserved environment. This generalises the linearly accelerated single electron case considered by Bell and Leinaas over 25 years ago. For the first time we have found analytic expressions for the $T_2$ (\ref{eqn:T2}), the concurrence (\ref{eqn:con3}) and the entanglement lifetime (\ref{eqn:t0}) of this system.\\ 
It is worth emphasising that while the timescales found here for the linear system prohibit any experimentation, there is a drastic improvement in the timescale when the accelerating electron is put into a circular orbit \cite{Bell:1982qr}.  Using the methods we have employed here, one could perform the calculation in the case when the motion is circular or when the acceleration in the IRF is simple harmonic. The latter situation is of interest as it would be a good theoretical basis in which to study the entanglement between spins attached to high frequency cantilevers.
\acknowledgements
The authors wish to thank G. Joshi, M. Schlosshauer and G. J. Milburn for useful discussions. JD was supported by the David Hay Memorial Fund and the Australian Research Council via its support for the Centre of Excellence for Mathematics and Statistics of Complex Systems. LCLH acknowledges the support of the Australian Research Council
(through the Centre of Excellence scheme, and an Australian Professorial
Fellowship DP0770715), the US National Security Agency (NSA), and the
Army Research Office (ARO) under contract number W911NF-08-1-0527.
\appendix \label{app:thomas}\section{Connection between Thomas precession and the infinitesimal Wigner rotation}
As discussed in the introduction a single boost $A(\vec{\bm{\beta}}+d\vec{\bm{\beta}})$ differs from the combination of boosts $A(\vec{\bm{\beta}})$ followed by $A(d\vec{\bm{\beta }}')$ by a rotation:
\begin{equation}
R(d\vec{\Omega})=A(\vec{\bm{\beta}}+d\vec{\bm{\beta}})[A(d\vec{\bm{\beta}}')A(\vec{\bm{\beta}})]^{-1},
\end{equation}
where $d\vec{\bm{\beta}}'$ is measured in the IRF and is related to $d\vec{\bm{\beta}}=\vec{\bm{v}}/c$ (see \cite{Jackson:1998}, chapter 11.8, \cite{JacksonNotation}) by:
\begin{equation}
d\vec{\bm{\beta}}'=\gamma d\vec{\bm{\beta}}+\frac{\gamma^3}{\gamma+1}d\vec{\bm{\beta}}\cdot\vec{\bm{\beta}}\vec{\bm{\beta}}.
\end{equation}
Using $R(d\vec{\Omega})A(d\vec{\bm{\beta}}')=A(d\vec{\bm{\beta}}')R(d\vec{\Omega})+\mathcal{O}(d\vec{\bm{\beta}})$ \cite{JacksonError} one finds:
\begin{eqnarray}
R(d\vec{\Omega})&=&A(-d\vec{\bm{\beta}}')A(\vec{\bm{\beta}}+d\vec{\bm{\beta}})A^{-1}(\vec{\bm{\beta}}),\\
&=&I-d\vec{\Omega}\cdot S,
\end{eqnarray}
where $S$ is the generator of rotations and 
\begin{equation}
d\vec{\Omega}=\frac{\gamma^2}{\gamma+1}\vec{\bm{\beta}}\times d\vec{\bm{\beta}}.
\end{equation}
One can verify that $-d\vec{\bm{\beta}}'$ is the three component vector of $A(\vec{\bm{\beta}}+d\vec{\bm{\beta}})\beta$, where $\beta=(\gamma,\gamma\vec{\bm{\beta}})$, in what follows we adopt the notation that $\vec{\beta_{\Lambda}}$ represents the three vector component of $\Lambda \beta$. Defining $L=A^{-1}$ and $\Lambda=A(\vec{\bm{\beta}}+d\vec{\bm{\beta}})$ we have,
\begin{equation}
R(d\vec{\Omega})=L^{-1}(\vec{\bm{\beta}}_{\Lambda})\Lambda L(\vec{\bm{\beta}}),
\end{equation}
which shows that the infinitesimal Thomas rotation is a special case of the Wigner rotation \cite{Milburn, PhysRevLett.89.270402} for the choice of $\Lambda$ given above. Therefore a state $|\vec{p} \lambda\rangle$ acted on by $U(\Lambda)$ (where $U$ is a two dimensional linear operator satisfying $U(\Lambda_1\Lambda_2)=U(\Lambda_1)U(\Lambda_2)$) can be written
\bea
U(\Lambda)|\vec{p} \lambda\rangle&=&U(\Lambda)U(L(\vec{\bm{\beta}}))|\vec{k} \lambda\rangle,\\
&=&U(L(\vec{\bm{\beta}}_{\Lambda}))U(L^{-1}(\vec{\bm{\beta}}_{\Lambda})\Lambda L(\vec{\bm{\beta}}))|\vec{k} \lambda\rangle,\\
&=&\sum_{\lambda'}D_{\lambda\lambda'}(\Lambda,\vec{p})|\vec{p}_{\Lambda} \lambda'\rangle.
\eea
where we have made use of $L(\vec{\bm{\beta}})^{\mu}_{\nu}k^{\nu}=p^{\mu}$, where $k$ is the rest frame four momentum $(mc^2,0)$. On the second line we have inserted the identity $U(L(\vec{\bm{\beta}}_{\Lambda}))U(L^{-1}(\vec{\bm{\beta}}_{\Lambda}))=1$ and identified the Wigner Rotation. The momentum dependent Wigner $D$-function is known to disentangle spin states \cite{PhysRevLett.89.270402} when either state has some spread in momentum. Putting the rotation into this form allows us to see that a continuous sequence of Wigner rotations caused by a constant acceleration in general leads to a mixing of the spin and momentum entanglement. We now complete the analysis by showing that this effect is just the Thomas precession, which can be accounted for by including an additional term in the Hamiltonian.\\
When the acceleration in the IRF is constant the moving frame rotates with a constant angular velocity,
\begin{equation}
\vec{\omega}_T=-\frac{d\vec{\Omega}}{dt}=\frac{\gamma^2}{\gamma+1}\frac{d\vec{\bm{\beta}}}{dt}\times\vec{\bm{\beta}}.
\end{equation}
Now we can write
\beq
U(R(d\vec{\Omega}))=1-\frac{d\vec{\Omega}}{dt}\cdot\mathbb{S}dt,
\eeq
where in our case we choose $\mathbb{S}$ to be the generators of the two dimensional representations of the unitary subgroup of $SL2(\mathbb{C}~)$, i.e., that satisfy the commutation relations:
\begin{align}
&[\mathbb{S}_i,\mathbb{S}_j]=~\epsilon_{ijk}\mathbb{S}_k,\\ 
&[\mathbb{S}_i,\mathbb{K}_j]=~\epsilon_{ijk}\mathbb{K}_k,\\
&[\mathbb{K}_i,\mathbb{K}_j]=-\epsilon_{ijk}\mathbb{S}_k,
\end{align}
where $\mathbb{K}$ ~are the hermitian subset of $SL2(\mathbb{C}~)$ ~corresponding to boosts. Since the Pauli matrices satisfy $[\sigma_i,\sigma_j]=2i\epsilon_{ijk}\sigma_k$ we can take,
\beq
\mathbb{S}_i=-\frac{i\sigma_i}{2},\quad \forall i\in \{x,y,z\}.
\eeq
Then 
\bea
U(R(d\vec{\Omega}))&=&1+\vec{\omega}_T\cdot\mathbb{S},\\
&=&1-\frac{i}{\hbar}\left(\frac{\hbar \vec{\omega}_T\cdot \vec{\sigma}}{2}\right)dt.
\eea
This equation is the infinitesimal form of the unitary time translation operator and thus the generator in brackets is the Thomas Hamiltonian,
\bea
H_{T}=\frac{\hbar\vec{\omega}_T\cdot\vec{\sigma}}{2}.
\eea
In the IRF the spin-field equation of motion is given by,
\beq
\frac{d\vec{s}}{d\tau}=\vec{\mu}\times B'
\eeq
where $B'$ is the field in the IRF. In the laboratory time,
\beq
\frac{d\vec{s}}{dt}=\gamma^{-1}\mu\times B'.
\eeq
Thus, the total Hamiltonian for the relativistic two component spinor can then be written,
\begin{equation}
H=-\gamma^{-1}\vec{\mu}\cdot B'+\frac{\hbar\vec{\omega}_T\cdot\vec{\sigma}}{2}.
\end{equation}


\begin{thebibliography}{999}
\bibitem{basischange}
This is just a change of basis of the standard polarization directions and should not be confused with the creation or annihilation operators which are obtained by separating the field into positive and negative frequencies.
\bibitem{photonrel}
These magnetic modes are related to photon modes through $B=\nabla \times A$.
\bibitem{justifyperturb}
Since the spin-field coupling strength is small a perturbative expansion is acceptable.
\bibitem{tracezero}
$\text{Tr}_B(V(\tau)\rho_B)=-\mu \sum\sigma_i\text{Tr}_B(B_i^{\dagger}\rho_B)$ and since $B_i=B^+_i+B_i^{-}$ where $B_i^{\pm}$ are the positive and negative frequencies, then as long as the field is not in a squeezed state (we assume $\rho_B$ is not) $\text{Tr}_B(B_i^{\pm}\rho_B)=0$.
\bibitem{epsilon}
Where the choice of $-i\epsilon$ has been made to recover the appropriate Greens function for this two-point correlation function (see \cite{BirrellandDavis} for a discussion on the analytic properties of Greens functions).
\bibitem{concurrenceclause}
In practice one could observe a loss of concurrence before this time, however, defining the timescale in this way is preferred since it is more meaningful to speak of the time taken for the spins to completely disentangle from each other as opposed to the time taken for them to evolve to some arbitrary value of concurrence.
\bibitem{zeroctime}
We note for completeness that the time taken for the concurrence to decay to the value $C$, is given by $\tau_C=\frac{3\pi}{8} \ln \left(\tfrac{3}{2 C +1}\right)\frac{\hbar c^6}{\mu^2a^3}.$
\bibitem{JacksonNotation}
However notice the slight difference in notation, $d\vec{\bm{\beta}}'\rightarrow\Delta\bm{\beta}$.
\end{thebibliography}
\end{document}